\documentclass[11pt]{article}

\usepackage{amssymb}
\usepackage{amsmath}
\usepackage{amsthm}

\usepackage{graphicx}
\usepackage{tikz}
\usetikzlibrary{arrows.meta}
\usetikzlibrary{automata, calc}
\usetikzlibrary{decorations.pathmorphing}
\tikzstyle{accepting}=[path picture={%
  \draw let 
    \p1 = (path picture bounding box.east),
    \p2 = (path picture bounding box.center)
    in
      (\p2) circle (\x1 - \x2 - 2pt);
  }]

\usepackage{xspace}
\usepackage{titlesec}
\titlespacing*{\subsection}{0pt}{3.25ex plus 1ex minus .2ex}{0pt}

\begin{document}




\title{Language Equivalence is Undecidable in VASS with Restricted Nondeterminism}

\author{Wojciech Czerwiński\thanks{Supported by the ERC grant INFSYS, agreement no. 950398} \\ University of Warsaw           \\ \texttt{wczerwin@mimuw.edu.pl} \and
Łukasz Orlikowski\thanks{Supported by the ERC grant INFSYS, agreement no. 950398}    \\ University of Warsaw           \\ \texttt{l.orlikowski@mimuw.edu.pl}}

\maketitle





\begin{abstract}
In this work, we extend undecidability of language equivalence for two-dimensional Vector Addition System with States (VASS) accepting by coverability condition. We show that the problem is undecidable even when one of the two-dimensional VASSs is deterministic and the other is history-deterministic. Moreover, we observe, that the languages of two history-deterministic VASSs are equal if and only if each can simulate the other. This observation allows us to extend the undecidability to any equivalence relation between two-sided simulation and language equivalence. 
\end{abstract}












\newtheorem{defi}{Definition}
\newtheorem{lemma}[defi]{Lemma}
\newtheorem{theorem}[defi]{Theorem}
\newtheorem{corollary}[defi]{Corollary}
\newtheorem{remark}[defi]{Remark}
\newtheorem{open}{Open problem}
\newtheorem{claim}[defi]{Claim}
\newcommand{\reals}[0]{\mathbb{R}}
\newcommand{\wctodo}[1]{\todo[inline, linecolor=blue, backgroundcolor=blue!30, bordercolor=black]{\textbf{Wojtek:} #1}}
\newcommand{\lotodo}[1]{\todo[inline, linecolor=yellow, backgroundcolor=yellow!30, bordercolor=black]{\textbf{{\L}ukasz:} #1}}
\newcommand{\N}[0]{\mathbb{N}}
\newcommand{\Z}[0]{\mathbb{Z}}
\newcommand{\support}[0] {
  \textup{support}}
\newcommand{\eff}[0] {
  \textup{eff}}
\newcommand{\maxdrop}[0] {
  \textup{maxdrop}}
\newcommand{\kdet}[0] {\textup{k-Det}}
\newcommand{\dett}[0] {\textup{Det}}
\newcommand{\hist}[0] {\textup{Hist}}
\newcommand{\khist}[0] {\textup{k-Hist}}
\newcommand{\kndet}[0] {\textup{k-NonDet}}
\newcommand{\ndet}[0] {\textup{NonDet}}
\newcommand{\unamb}[0] {\textup{Unamb}}
\newcommand{\kunamb}[0] {\textup{k-Unamb}}
\newcommand{\xdet}[1]{\textup{#1-}\dett}
\newcommand{\xunamb}[1]{\textup{#1-}\unamb}
\newcommand{\xndet}[1]{\textup{#1-}\ndet}
\newcommand{\xhist}[1]{\textup{#1-}\hist}
\newcommand{\lang}{\textup{Lang}}
\newcommand{\twosim}{\textup{Two-Sided-Sim}}
\newcommand{\naturalrel}{$\twosim \subseteq R \subseteq \lang$\xspace}
\newcommand{\set}[1]{\{#1\}}
\newcommand{\proofnewline}{$ $\newline}
\newenvironment{proofnodot}[1][\proofname]{%
  \par\pushQED{\qed}\normalfont
  \topsep6\p@\@plus6\p@\relax
  \trivlist
  \item[\hskip\labelsep
        \itshape
        #1\@addpunct{ }
  ]%
}{\popQED\endtrivlist\@endpefalse}
\makeatother

\section{Introduction}
Vector Addition System (VAS) is a central model in concurrency theory, which has a lot of applications both in theory and in practical modeling~\cite[Section 5]{DBLP:journals/siglog/Schmitz16}. A $d$-dimensional VASS ($d$-VASS) is a finite automaton equipped additionally with $d$ nonnegative integer counters that are updated by transitions, in such a way that the counter values cannot drop below
zero. VASSs have no capability to zero-test counters and hence the model is not Turing-complete. Nowadays, there is an active line of research on VASs and almost equivalent models such as Petri Nets and Vector Addition Systems with States (VASSs). For instance, recently the complexity of the reachability problem for VASSs was settled to be Ackermann-complete \cite{Leroux21, ackermann_WCZLO, DBLP:conf/lics/LerouxS19}.

Languages of VASSs have been already studied for several years including such problems as regularity \cite{regular_petri} or universality \cite{universality_cover} of a language of a VASS, as well as other problems such as whether there exists VASS recognising the same language with some properties (e.g. being deterministic) \cite{history-deterministic, deterministic, our-unambiguous}. For the languages of VASSs, two different acceptance conditions have been studied: reachability and coverability. Most work focuses on coverability acceptance, because for reachability the universality problem is already undecidable (folkrole; see, for instance, \cite{our-unambiguous}), whereas for coverability, universality is decidable using techniques based on well quasi-orders \cite{universality_cover}. One of the most natural questions for any pair of systems, that is language equivalence problem, asking whether languages of two given systems are equal, has also been studied for VASSs. It is undecidable for the class of all VASS as proven first in 1975 by Araki and Kasami \cite{Araki_Kasami} for reachability acceptance and later by Jančar \cite{jancar_first}, whose construction also works for acceptance by coverability condition and proves undecidability already for two-dimensional VASSs. Araki and Kasami also proved that the language equivalence problem for deterministic VASSs can be reduced to the reachability problem, which was shown to be decidable in 1981 by Mayr \cite{Mayr81}. On the other hand, Jančar generalised his construction to any equivalence relation between isomorphism and language equivalence \cite{jancar_second}, including many interesting equivalences like bisimulation, two-sided simulation or multiplicity equivalence. These works left as an open problem the decidability of language equivalence for VASSs accepting by coverability condition in dimension one (i.e. with only one counter). This problem was finally proven to be undecidable by Hofman, Mayr and Totzke in 2013 \cite{HMT13}.

Because the language equivalence problem and also many other equivalence problems are undecidable for general VASSs \cite{jancar_first, jancar_second} there is a line of research in the literature to study language equivalence for subclasses of VASSs with restricted nondeterminism. This is mostly inspired by the fact, that by results of Araki, Kasami and Mayr \cite{Araki_Kasami, Mayr81} language equivalence is decidable for deterministic VASSs. Moreover, recently there has been a lot of research on unambiguous systems (in which for each accepted word there is exactly one accepting run). For instance in 2022 language equivalence for unambiguous VASSs accepting by states was shown to be decidable by Czerwiński and Hofman \cite{concur-unambigous}. 

Another restriction of nondeterminism for VASSs studied in the literature is history-determinism \cite{history_OCN, history-deterministic}. Several motivations have driven the introduction of history-determinism. Henzinger and Piterman \cite{games} introduced it for solving games without determinisation, Colcombet \cite{cost_functions} for cost functions, and Kupferman, Safra, and Vardi \cite{word_automata} for recognising derived tree languages of word automata. A nondeterministic automaton is called history-deterministic when its nondeterministic choices can be resolved step by step by a resolver. For any input word, as each letter is read, a resolver decides which transition to take next. Following the resolver must preserve the automaton’s language, meaning it accepts exactly the same words as the original. For history-deterministic VASSs it was shown in \cite{history-deterministic}, that problems such as language inclusion for two history-deterministic VASSs, checking whether given VASS is history-deterministic and checking whether given VASS has equivalent history-deterministic one is undecidable. On the other hand language inclusion is decidable if history-deterministic VASSs are one-dimensional \cite{history_OCN}.

\subsection*{Our contribution}
In this work, we study the decidability of being in equivalence relations for Vector Addition Systems with States accepting by coverability condition. We follow the line of research of studying language equivalence problems for VASSs with restricted nondeterminism \cite{Araki_Kasami, concur-unambigous}. Contrary to the previous works, we study the non-symmetric case when one VASS is deterministic and the second one is nondeterministic. We revisit the construction of Jančar \cite{jancar_first, jancar_second} showing that the language equivalence problem for VASS is undecidable and observe, that it is undecidable even if one VASS is deterministic and the second one is history-deterministic. Further, we observe that the languages of two history-deterministic VASSs are equal if and only if each can simulate the other. This observation allows us to extend our undecidability result to any equivalence relation between two-sided simulation and language equivalence, as stated in Theorem~\ref{thm:equivalence}. 

\subsection*{Organisation of the paper}
We begin by introducing preliminary notions in Section \ref{sec:prelim}. Section \ref{sec:equivalence} is devoted to proving our two main theorems, namely Theorem \ref{thm:language_equivalence} and Theorem \ref{thm:equivalence}. Finally, in Section \ref{sec:future} we discuss interesting
future research directions.

\section{Preliminaries}\label{sec:prelim}
\subsection*{Basic notions}
For $a,b \in \N$ such that $a \leq b$ we write $[a, b]$ for the set of integers $\{a, a+1, \ldots, b\}$.
For $a \in \N$ we write $[a]$ for the set $[1, a]$. For $a \in \Z$ we denote by $\vec{a}$ the constant vector $(a,a, \ldots, a) \in \Z^d$. For a vector $v \in \Z^d$ we write $v_i$ for the $i$-th entry of $v$. For two vectors $v, u \in \N^d$ we write $v \geq u$ if for all $i \in [d]$ we have $v_i \geq u_i$. For a word $w$ we denote by $|w|$ the number of letters in $w$.
\subsection*{Vector Addition Systems with States}
A $k$-dimensional Vector Addition System with States ($k$-VASS) is a nondeterministic finite automaton  with $k$
non-negative integer counters. Transitions of the VASS manipulate these counters. Formally, we say, that a VASS $V$ consists of a finite alphabet $\Sigma$, a finite set of automaton states $Q$, an initial
configuration $c_0 \in Q \times \N^k$, a finite set of final configurations $F \subseteq Q \times \N^k$ and a transition relation $\delta \subseteq Q \times \Sigma \times \Z^k \times Q$. For a transition $t = (s, a, v, s') \in \delta$ we say, that it is over the letter $a$, or equivalently, it reads the letter $a$. We also write $\eff(t)$ for the effect of the transition $t$, which is $v$. 

A $k$-VASS can be seen as an infinite-state labelled transition system in which each configuration is a pair $(s, u) \in Q \times \N^k$. We denote such configuration as $s(u)$.  A transition $t = (s, a, v, s') \in \delta$ can be fired in a configuration $q(u)$ if and only if $q=s$ and $u' \geq 0$ where $u'=u+v$. After firing the transition the configuration is changed to $s'(u')$. We also define run as a sequence of transitions, which can be fired one after another from some configuration. For a run $\rho=t_1t_2\ldots t_n$ we write $\eff(\rho)$ for $\Sigma_{i=1}^n \eff(t_i)$. If we want to say something only about $j$ (for $j \in [k]$) entry of the $\eff(\rho)$ we write $\eff_j(\rho)$. We say that run $\rho$ is from configuration $q(u)$ to $p(u')$ if the sequence of transitions can be fired from $q(u)$ and the final configuration is $p(u')$. We say, that a run $\rho=t_1\ldots t_n$ is over $w=\lambda_1\ldots \lambda_n \in \Sigma^*$ (or reads $w$) if for each $i \in [n]$ transition $t_i$ is over $\lambda_i$. We say that the length of a run $\rho$ is equal to $n$ if it consists of $n$ transitions. For the length of the run, we write $|\rho|$. We say, that a word $w$ is read from configuration $q(u)$ if there exists run $\rho$ over $w$ from configuration $q(u)$.

\subsection*{Languages of VASSs}
In our work, we consider only coverability languages and their special case, trace languages. A run of a VASS is accepting if it starts in an initial configuration $c_0 \in I$ and ends in an accepting configuration. 
A configuration is accepting if it covers some configuration from the finite set of final configurations $F$.
In other words configuration $q(v)$ is accepting if there exists a configuration $q(v') \in F$ such that $v \geq v'$.
For a VASS $V$ we define its language as the set of all words read by accepting runs and we denote it by $L(V)$. Languages defined this way are called coverability languages. We often refer to trace languages of VASSs. It is a special case of coverability languages where the set of final configurations $F$ is equal to $Q \times \{\vec{0}\}$, namely all configurations are accepting. For a VASS $V$ we denote its trace language as $L_T(V)$. 

Notice, that one can also consider reachability languages, where a configuration is accepting only if it belongs to the set of final configurations. 

We say, that VASS $V$ is deterministic if and only if for each state $q$ of the VASS and each letter $a \in \Sigma$, where $\Sigma$ is an alphabet of the VASS, we have at most one transition $t$ leaving $q$ such that it is over $a$.

We say, that VASS $V$ is unambiguous if and only if for every $w \in L(V)$ we have exactly one accepting run over $w$. Please note, that deterministic VASS is a special case of unambiguous VASS, but unambiguous VASS does not have to be deterministic. We say that language is unambiguous if and only if it can be recognised by an unambiguous VASS and ambiguous if and only if an unambiguous VASS can not recognise it.

For a $k$-VASS consider a function $r : (Q \times \N^k \times \delta)^* \times (Q \times \N^k) \times \Sigma \rightarrow \delta$ that, given a history of the run (configurations and taken transitions), current configuration $q(v)$ and a next letter $\lambda \in \Sigma$,
returns a transition over $\lambda$, which can be fired from $q(v)$. 
Let us call $r$ a resolver. We say, that $k$-VASS $V$ is history-deterministic if and only if it has one initial configuration and there exists a resolver $r$ such that for each $w \in L(V)$ run $\rho$ over $w$ from the initial configuration given by the resolver is accepting.

\section{Deciding equivalence relations}\label{sec:equivalence}
In this section, we revisit the famous result by Jančar \cite{jancar_second} about the undecidability of a wide range of equivalence relations for VASSs. Jančar showed that for any equivalence relation between isomorphism and language equivalence, the question of whether two VASSs are equivalent is undecidable. The proof is based on a reduction from an undecidable problem - the reachability problem for two-counter machines \cite{minsky}. For a two-counter machine $M$ Jančar constructed two $2$-VASSs $A$ and $B$ such that $A$ and $B$ are equivalent if and only if the machine $M$ does not reach halting configuration. In Remark~3.4 he describes a simpler, asymmetric construction \cite[Remark~3.4]{jancar_first}. We follow this remark and observe, that the modified construction produces one deterministic VASS and one history-deterministic VASS. To the best of our knowledge, this was never stated before. Similarly, also decidability of language equivalence for the case when one VASS is deterministic and the other one is nondeterministic was unknown. We formally state our result in Theorem \ref{thm:language_equivalence}. Further, we observe, that languages of history-deterministic VASSs $A$ and $B$ are equal if and only if $A$ can simulate $B$ and vice versa. Therefore, we extend this result to more equivalence relations. The extension is formally stated in Theorem \ref{thm:equivalence}. 

As we said, the construction of Jančar is based on the reduction from the reachability problem for two-counter machines, which is undecidable \cite{minsky}. We recall the notion of two-counter machines in a syntax, which is suitable for us.
    A \emph{two-counter machine (2CM)} $M=(Q, q_i, q_f, \delta)$ consists of a finite set of states $Q$, initial state $q_i \in Q$, final state $q_f \in Q$ and a finite set of transitions $\delta \subseteq Q \times \Gamma \times Q$, where $\Gamma = \{inc_1, inc_2, dec_1, dec_2, z_1, z_2\}$ are operations on the counters. Moreover, there are no outgoing transitions from $q_f$ and for every state $q \in Q$ such that $q \neq q_f$ we have one of the following options:
    \begin{enumerate}
        \item There is exactly one transition leaving $q$ of the form $(q, inc_i, q')$ for some $q' \in Q$. This transition goes from $q$ to $q'$ and increments counter $i$ by one.
        \item There are exactly two transitions leaving $q$ of the form $(q, dec_i, q_1)$ and $(q,z_i, q_2)$ for some $q_1, q_2 \in Q$. The first transition goes from $q$ to $q_1$ and decrements counter $i$ while the second transition goes from $q$ to $q_2$ and performs zero-test on the $i$ counter, that means it can only be fired when counter $i$ is zero.
    \end{enumerate}
As we said the reachability problem, that means whether 2CM can reach some configuration of the form $q_f(x,y)$ for $x,y \in \N$ from $q_i(0, 0)$ is undecidable. Now we are ready to formulate and prove one of the main results of this work, that it is undecidable to determine whether for two $2$-VASSs, one deterministic and the second one history-deterministic, their trace languages are equivalent.
\begin{theorem}\label{thm:language_equivalence}
    It is undecidable to determine whether a given deterministic $2$-VASS and a given history-deterministic $2$-VASS have equal trace languages.
\end{theorem}
\begin{proof}
    We proceed by a reduction from the reachability problem for 2CM. For a 2CM $M$ we construct a deterministic VASS $A$ and history-deterministic VASS $B$ such, that:
    \begin{enumerate}
        \item If $M$ halts starting from zero vector then there exists $w \in L_T(A)$ such that $w \notin L_T(B)$. Hence trace languages of $A$ and $B$ are not equal.\label{cond:first_lang}
        \item If $M$ does not halt starting from zero vector then $L_T(A) = L_T(B)$. \label{cond:second_lang}
    \end{enumerate}
    Therefore $M$ halts if and only if $L_T(A) \neq L_T(B)$. Hence in order to show the undecidability of trace language equivalence it is enough to show the construction and prove the above properties. In the proof we first describe the construction of $A$ and $B$. Then we show that they fulfil condition \ref{cond:first_lang} and condition \ref{cond:second_lang}. Finally, we show that $A$ is deterministic and $B$ is history-deterministic.
    \subsection*{Construction}
            \begin{figure}
    \centering
    \begin{tikzpicture}
        \begin{scope}[initial text = \textcolor{red}{}]
            \node[] (A) at (0,0) {$q_i$};
            \node[] (B) at (0,-2) {$q_1$};
            \node[] (C) at (0, -4) {$q_f$};
            \node[] (D) at (0,1) {2CM $M$};
        \end{scope}
        
        \begin{scope}[>={Stealth[black]},
                      every edge/.style={draw=black,very thick}]
            \draw[] (-2, 1.5) rectangle (2, -4.5);
            \path [->] (A) edge [bend left] node [right] {$inc_1$} (B);
            \path [->] (B) edge [right] node {$z_2$} (C);
            \path [->] (B) edge [bend left] node [left] {$dec_2$} (A);
        \end{scope}
    \end{tikzpicture}
    \begin{tikzpicture}
        \begin{scope}[initial text = \textcolor{red}{}]
            \node[] (A) at (0,0) {$q_i$};
            \node[] (B) at (0,-2) {$q_1$};
            \node[] (E) at (-1.5, -3) {$q_2$};
            \node[] (C) at (0, -4) {$q_f$};
            \node[] (D) at (0,1) {$2$-VASS $N$};
        \end{scope}
        
        \begin{scope}[>={Stealth[black]},
                      every edge/.style={draw=black,very thick}]
            \draw[] (-2, 1.5) rectangle (2, -4.5);
            \path [->] (A) edge [bend left] node [right] {$inc_1$} (B);
            \path [->] (E) edge [above] node {$z_2$} (C);
            \path [->] (B) edge [above] node {$z_2$} (E);
            \path [->] (B) edge [bend left] node [left] {$dec_2$} (A);
        \end{scope}
    \end{tikzpicture}
    \begin{tikzpicture}
        \begin{scope}[initial text = \textcolor{red}{}]
            \node[] (A) at (0,0) {$q_{i_A}$};
            \node[] (B) at (0,-2) {$q_{1_A}$};
            \node[] (E) at (-1.5, -3) {$q_{2_A}$};
            \node[] (C) at (0, -4) {$q_{f_A}$};
            \node[] (D) at (0,1) {$2$-VASS $A$};
            \node[] (F) at (1.5, -3) {$q_h$};
        \end{scope}
        
        \begin{scope}[>={Stealth[black]},
                      every edge/.style={draw=black,very thick}]
            \draw[] (-2, 1.5) rectangle (2, -4.5);
            \path [->] (A) edge [bend left] node [right] {$inc_1$} (B);
            \path [->] (E) edge [above] node {$z_2$} (C);
            \path [->] (B) edge [above] node {$z_2$} (E);
            \path [->] (B) edge [bend left] node [left] {$dec_2$} (A);
            \path [->] (C) edge [below] node {$h$} (F);
        \end{scope}
    \end{tikzpicture}
    \begin{tikzpicture}
        \begin{scope}[initial text = \textcolor{red}{}]
            \node[] (A) at (0,0) {$q_{i_A}^B$};
            \node[] (B) at (0,-2) {$q_{1_A}^B$};
            \node[] (E) at (-1.5, -3) {$q_{2_A}^B$};
            \node[] (C) at (0, -4) {$q_{f_A}^B$};
            \node[] (D) at (1.5,1) {$2$-VASS $B$};
            \node[] (F) at (1.5, -3) {$q_h^B$};
            \node[] (G) at (3,0) {$q_{i_B}$};
            \node[] (H) at (3,-2) {$q_{1_B}$};
            \node[] (I) at (4.5, -3) {$q_{2_B}$};
            \node[] (J) at (3, -4) {$q_{f_B}$};
            \node[] (K) at (1.5, -2) {$q_{1_B}'$};
        \end{scope}
        
        \begin{scope}[>={Stealth[black]},
                      every edge/.style={draw=black,very thick}]
            \draw[] (-2, 1.5) rectangle (5, -4.5);
            \path [->] (A) edge [bend left] node [right] {$inc_1$} (B);
            \path [->] (E) edge [above] node {$z_2$} (C);
            \path [->] (B) edge [above] node {$z_2$} (E);
            \path [->] (B) edge [bend left] node [left] {$dec_2$} (A);
            \path [->] (C) edge [below] node {$h$} (F);
            \path [->] (G) edge [bend left] node [right] {$inc_1$} (H);
            \path [->] (H) edge [above] node {$z_2$} (I);
            \path [->] (I) edge [above] node {$z_2$} (J);
            \path [->] (H) edge [below] node {$z_2$} (K);
            \path [->] (K) edge [left] node {$z_2$} (C);
            \path [->] (H) edge [bend left] node [left] {$dec_2$} (G);
        \end{scope}
    \end{tikzpicture}
\caption{Example of the construction of $2$-VASSs $A$ and $B$ and auxiliary $2$-VASS $N$ from 2CM $M$. Effects on the counters corresponds to letters except from two transitions: the first transition from $q_{1_B}$ to $q_{1_B}'$ which decrements second counter by $1$ and the second transition from $q_{1_B}'$ to $q_{f_A}^B$ which increments second counter by $1$.}\label{img:construction}
\end{figure}
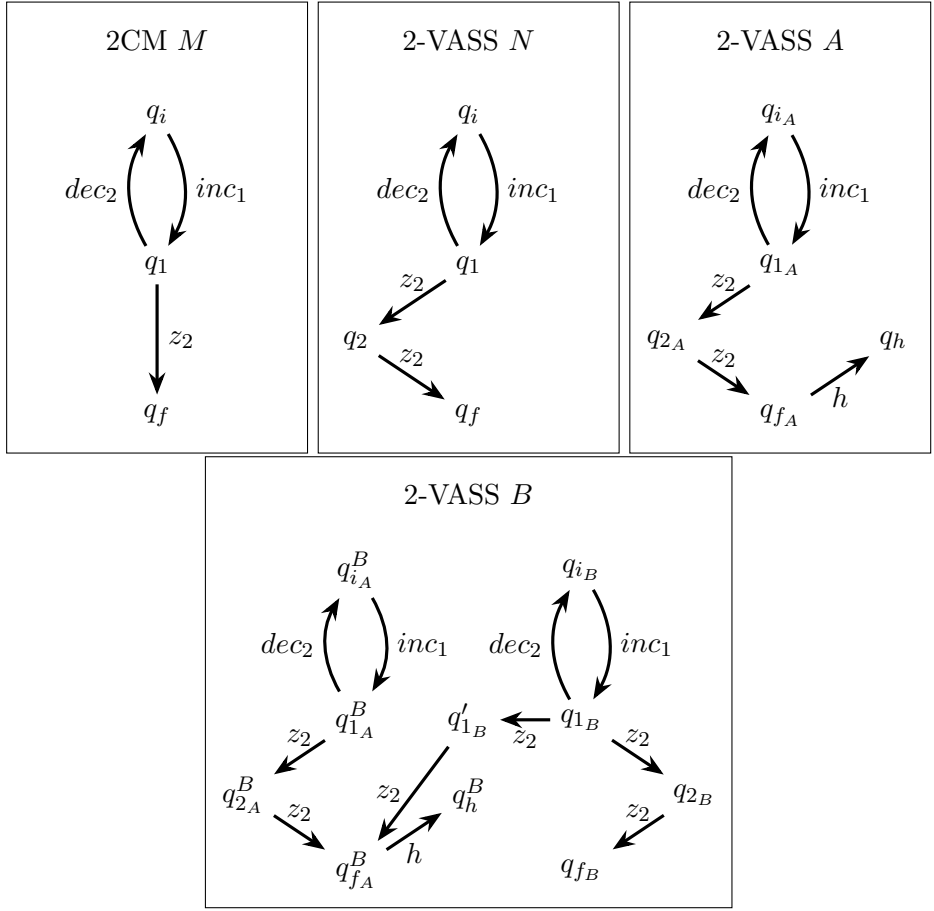
    Now we describe how to construct $2$-VASSs $A$ and $B$ from the 2CM $M$. An example of construction is presented in Figure \ref{img:construction}. First, we construct an auxiliary $2$-VASS $N$. Intuitively speaking behaviour of VASS $N$ should correspond to the behaviour of the two-counter machine $M$. The states of $N$ are the states of $M$ and some number of additional states, which we create later. For each transition $t$ of $M$ from state $q$ to $q'$ performing action $\lambda \in \{inc_1, inc_2, dec_1, dec_2\}$ we add a transition $t'$ from $q$ to $q'$ in $N$ over letter $\lambda$ having the same effect on the counters as $t$. Finally for each transition $t$ of $M$ from $q$ to $q'$ performing action $\lambda \in \{z_1, z_2\}$ we create a state $q_t$ in $N$ and two transitions, both over $\lambda$. The first one goes from $q$ to $q_t$ and the second one from $q_t$ to $q'$. These transitions have no effect on the counters.

    Now we can construct VASS $A$. We start the construction by taking VASS $N$ and adding an additional state $q_h$. We also add transition $t$ from the state corresponding to the halting state of $M$ to the state $q_h$ with no effect on the counters and over letter $h$. Hence we constructed VASS $A$ over alphabet $\Sigma=\{inc_1, inc_2, dec_1, dec_2, z_1, z_2, h\}$. For every state $q$ of $N$ we denote its copy in $A$ as $q_A$.

    Now we are ready to construct VASS $B$. Intuitively speaking $B$ is a union of $A$ and $N$. One can go from $N$ to  $A$ by taking two transitions each over $z_i$. Intuitively these transitions perform a cheated (on a non-zero counter) zero-test on the counter $i$. In order to ensure that we really perform a cheated zero-test we decrement and then increment the counter by one. Formally, we begin the construction by taking a disjoint union of VASS $N$ and VASS $A$. For every state $q$ of $N$ we denote its copy in $B$ as $q_B$ and for each state $q_A$ of $A$ we denote its copy in $B$ as $q_A^B$. Finally, we denote copy of the state $q_h$ from $A$ in $B$ as $q_h^B$. Let $q$ be a state of $N$ corresponding to a state of $M$, such that there exists transition $t$ leaving $q$ via letter $z_i$ to another state $q_t$. Then there exists exactly one state $p$ in $N$ such that from $q_t$ there is a transition over $z_i$ to $p$. For each such triple $(q, q_t, p)$ we add a state $q_B'$ in $B$ and transitions, both over $z_i$ from $q_B$ to $q_B'$ and from $q_B'$ to $p_A^B$. The first decrementing the $i$-th counter by one and the second one incrementing the $i$-th counter by one.

    Finally, we have to set initial configurations of $A$ and $B$. Let $q$ be the state of $N$ corresponding to the initial state of 2CM $M$. We set initial configuration of $A$ to be $q_A(0, 0)$ and initial configuration of $B$ to be $q_B(0, 0)$.

    The idea of the above construction is that trace languages of $A$ and $B$ are different if and only if $M$ halts. In particular $A$ can read the word corresponding to the halting run and $B$ cannot.
    \subsection*{First condition}
    If $M$ halts it is easy to show, that there exists $w \in L_T(A)$ such that $w \notin L_T(B)$. We can look at the halting run of $M$ (recall that two-counter machines are deterministic, hence $M$ has exactly one run). Let the halting sequence of operations be: $\lambda_1\lambda_2\ldots\lambda_n$ then we can create $w$ by duplicating each $\lambda_i$ such that $\lambda_i \in \{z_1, z_2\}$ and adding letter $h$ at the end.
    It is clear, that $w \in L_T(A)$ because $A$ is constructed as a copy of $M$ with letters on transitions corresponding to operations of $M$ and each transition of $M$ with $z_i$ operation split into two transitions over $z_i$. Finally, the letter $h$ can be read, since after reading the preceding letter, VASS $A$ reaches a state that was constructed as a copy of the halting state of $M$.

    On the other hand $w \notin L_T(B)$. Assume, towards contradiction, that $w \in L_T(B)$. Recall, that $B$ was constructed as a union of a copy of $A$ and a copy of $N$ with some additional transitions. Since we are starting in the part of $B$, which was constructed as a copy of $N$ and the only transition over the letter $h$ is in the part of $B$, which was constructed as a copy of $A$ in order to accept $w$ we have to, at some point cross from the part corresponding to $N$ to the part corresponding to $A$. This is only possible by executing at some point sequence $z_iz_i$ with non-zero counter $i$ before. However, this is not possible as $w$ is constructed from the correct run of $M$ and when we are reading infix $z_iz_i$ then the $i$-th counter can be only zero and therefore crossing with transitions performing on the $i$-th counter first $-1$ and then $+1$ is not possible.

    \subsection*{Second condition}

    We will show, that if $M$ does not halt then $L_T(A) = L_T(B)$. We proceed in two steps. First, we show that $L_T(B) \subseteq L_T(A)$ and later that $L_T(A) \subseteq L_T(B)$. Actually the inclusion $L_T(B) \subseteq L_T(A)$ always holds and inclusion $L_T(A) \subseteq L_T(B)$ holds if and only if $M$ does not halt.

    To show $L_T(B) \subseteq L_T(A)$ we will need the following, more general lemma, intuitively saying that if some word can be read by $B$ (starting from the part constructed as copy of $N$) then it can be also read from some of the configurations of $A$.
    \begin{lemma}\label{lem:A_sim_B}
        For each state $q$ of $N$, each vector $u \in \N^2$ and each word $w$ if one can read word $w$ from the configuration $q_B(u)$ in $B$ then one can read word $w$ from the configuration $q_A(u)$ in $A$.
    \end{lemma}
    \begin{proof}
      We prove this by induction on the length of the word $w$. The case when $|w|=0$ is obvious and when $|w| = 1$ observe, that from $q_B(u)$ we can not read the letter h. Because $w$ is read from $q_B(u)$ we have that $w \in \{inc_1, inc_2, dec_1, dec_2, z_1, z_2\}$. Observe, from the construction of $B$, that there exist a transition $t$ over $w$ in $B$, which goes from state $q_B$ to state $p_B$. Observe, that on the other hand, from the construction of $A$ there exist a transition $t'$ over $w$ in $A$, which goes from the state $q_a$ to the state $p_A$ and has the same effect as $t$. Therefore we can also read $w$ from $q_A(u)$.

      Now we have to show the induction step. Let us take some $w$ such that $|w| \geq 2$ and assume Lemma \ref{lem:A_sim_B} for all $v$ such that $|v| < |w|$. Let $\rho$ be a run reading $w$ from $q_B(u)$. Let us look at the first transition $t$ of $\rho$. We have two cases. The first option is that transition $t$ goes to a configuration $p_B(u')$. From construction of $A$ and $B$ we have transition $t'$ over the same letter and having the same effect as $t$ from $q_A$ to $p_A$. Therefore we can fire transition $t'$ from $q_A$ and go to configuration $p_A(u')$. Now it is enough to apply the induction assumption to the word $w$ without the first letter, vector $u'$ and state $p$ of $N$.
      
        The second option is that after firing of the two first transitions of $\rho$, that is transitions $t_1$ and $t_2$ from $q_B(u)$ we go to configuration $p_A^B(u')$. Observe, that then we have $u'=u$ and word read by $t_1t_2$ is equal to $z_iz_i$ for $i \in \set{1,2}$. Moreover, from construction of $A$ and $B$ we have transitions $t_1'$ and $t_2'$ in $A$ such that:
        \begin{itemize}
            \item We can fire $t_1t_2$ from $q_A(u)$ and go to $p_A(u)$
            \item Word read by $t_1t_2$ is the same as the word read by $t_1't_2'$, namely $z_iz_i$ for $i \in \set{1,2}$
        \end{itemize}
        Let $v$ be $w$ without first two letters. Observe that states reachable from $p_A^B$ are only copies of the states of $A$. Hence if we can read $v$ from $p_A^B(u)$ then we can read $v$ from $p_A(u)$. Therefore we conclude that we can read $w$ from $q_A(u)$ also in this case.
    \end{proof}
    Observe, that $L_T(B)$ and $L_T(A)$ are the sets of words, which can be read from $q_B(0,0)$ and $q_A(0,0)$ respectively. Thus by applying Lemma \ref{lem:A_sim_B} to state $q$ of $N$, vector $(0,0)$ and every $w \in L_T(B)$ we conclude that $L_T(B) \subseteq L_T(A)$.
    Now we have to show, that $L_T(A) \subseteq L_T(B)$. Therefore let us take $w \in L_T(A)$ and show that $w \in L_T(B)$. 
    
    The first case is that $w$ does not contain the letter h. Recall, that $A$ has only one transition over the letter h and all the other transitions come from $N$. Therefore to accept $w$ VASS $A$ uses only transitions and states coming from $N$. Therefore we can use only the part of $B$, which was created as a copy of $N$ to accept $w$. Hence $w \in L_T(B)$. 
    
    The second case is that word $w$ contains letter h. Situation in this case is depicted in Figure \ref{img:help_1}. Because there is only one transition reading h in $A$ and no transition leaves the state to which this transition leads the letter $h$ is the very last letter of $w$ and all the other letters are different. Recall, that $M$ does not halt. Therefore at some point, $w$ differs from the run of machine $M$. Because of how $A$ was constructed, the only possibility for the word $w$ to differ from the run of $M$ is to read sequence $z_iz_i$ when i-th counter is not zero. Hence, we can decompose $w$ as $w = uz_iz_iv$ for some words $u,v$ and $i \in \{1,2\}$ such that $\eff_i(\rho) \neq 0$ where $\rho$ is the part of the run reading $u$. Observe, that $\rho$ contains only transitions coming from $N$, because the only other transition is over h. Therefore we can also read $u$ from the initial configuration of $B$, because we start in part of $B$, which was created as a copy of $N$.  Moreover, if $A$ is in configuration $p_A(x)$ after reading $u$ we have that $B$ is in configuration $p_B(x)$. After reading $z_iz_i$ from $p_A(x)$ in $A$ we go to the configuration $r_A(x)$. On the other hand, because $x_i > 0$, observe that from $p_B(x)$ in $B$ we can fire transitions over $z_iz_i$ and go to configuration $r_A^B(x)$. Therefore from the configuration $r_A^B(x)$ in $B$ we also can read word $v$ and therefore accept word $w$. Hence $w \in L_T(B)$.


        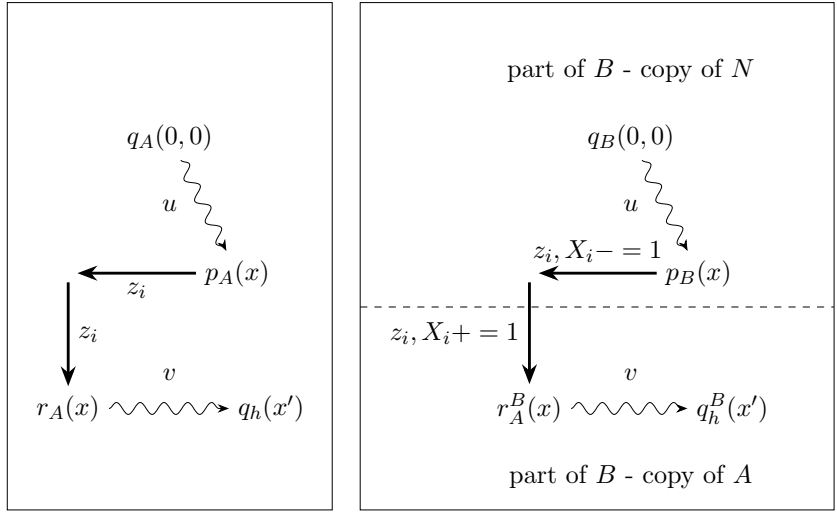
\begin{figure}
    \centering
    \scalebox{0.9}{
    \begin{tikzpicture}
        \begin{scope}[initial text = \textcolor{red}{}]
            \node[] (A) at (0,-1) {$q_A(0,0)$};
            \node[] (B) at (1,-3) {$p_A(x)$};
            \node[] (C) at (-1.5, -3) {};
            \node[] (D) at (-1.5,-5) {$r_A(x)$};
            \node[] (E) at (0, -2) {$u$};
            \node[] (F) at (1.5, -5) {$q_h(x')$};
            \node[] (G) at (0, -4.5) {$v$};
        \end{scope}
        
        \begin{scope}[>={Stealth[black]},
                      every edge/.style={draw=black,very thick}]
            \draw[] (-2.4, 1) rectangle (2.4, -6.5);
            \draw [->,decorate,decoration={snake}] (A) -- (B);
            \path [->] (B) edge [below] node {$z_i$} (C);
            \path [->] (C) edge [right] node {$z_i$} (D);
            \draw [->,decorate,decoration={snake}] (D) -- (F);
        \end{scope}
    \end{tikzpicture}
    }
    \scalebox{0.9}{
    \begin{tikzpicture}
        \begin{scope}[initial text = \textcolor{red}{}]
            \node[] (A) at (0,-1) {$q_B(0,0)$};
            \node[] (B) at (1,-3) {$p_B(x)$};
            \node[] (C) at (-1.5, -3) {};
            \node[] (D) at (-1.5, -5) {$r_A^B(x)$};
            \node[] (E) at (0, -2) {$u$};
            \node[] (F) at (0, 0) {part of $B$ - copy of $N$};
            \node[] (G) at (0, -6) {part of $B$ - copy of $A$};
            \node[] (H) at (1.5, -5) {$q_h^B(x')$};
            \node[] (I) at (0, -4.5) {$v$};
        \end{scope}
        
        \begin{scope}[>={Stealth[black]},
                      every edge/.style={draw=black,very thick}]
            \draw[] (-4, 1) rectangle (3, -6.5);
            \draw[dashed] (-4, -3.5) -- (3, -3.5);
            \draw [->,decorate,decoration={snake}] (A) -- (B);
            \draw [->,decorate,decoration={snake}] (D) -- (H);
            \path [->] (B) edge [above] node {$z_i, X_i-=1$} (C);
            \path [->] (C) edge [left] node {$z_i, X_i+=1$} (D);
        \end{scope}
    \end{tikzpicture}
    }
\caption{Situation in $A$ on the left and situation in $B$ on the right.  $X_i -=1$ and $X_i+=1$ means respectively decrementing and incrementing $i$-th counter by one.}\label{img:help_1}
\end{figure}
    \subsection*{$A$ is deterministic}
    VASS $A$ is deterministic because for each state $q$ and each $\lambda \in \Sigma$ there is at most one transition leaving $q$ over $\lambda$.
    \subsection*{$B$ is history-deterministic}
    In VASS $B$ we have three types of states: coming from $A$, coming from $N$ and added in the construction of $B$. We have a nondeterministic choice only in states coming from $N$ in which we have exactly two transitions over $z_i$ from which we have to choose one to resolve nondeterminism. If the value of the i-th counter is zero, we do not have a choice, because one of these transitions decrements counter $i$, so we have to take the other one. In all of the other cases, we have two transitions available. To resolve nondeterminism we will take the one decrementing counter $i$. 
    
    The situation is depicted in Figure \ref{img:help_2}. Let $q_B'(c_1)$ be a configuration in which we will be after choosing the transition decrementing counter $i$ and let $p_B(c_2)$ be a configuration in which we will be after choosing the second transition. To show, that we resolved nondeterminism correctly we have to show, that if one can read the word $w$ from configuration $p_B(c_2)$ then $w$ can be read also from $q_B'(c_1)$. Let us see, that from both configurations, we have exactly one option: a transition over $z_i$. The transition from $q_B'(c_1)$ increments $i$-th counter while the transition from $p_B(c_2)$ have no effect on the counters. Hence after reading the next letter, we will be in configurations $r_A^B(c_2)$ and $r_B(c_2)$ with equal values of the counters. Observe that if a word $v$ is read from $r_A^B(c_2)$ the run uses only transitions of $A$. Therefore, it is enough to show that if $v$ can be read from $r_B(c_2)$ in B then it can also be read from $r_A(c_2)$ in $A$. For this we apply Lemma \ref{lem:A_sim_B} to state $r$ of $N$, vector $c_2$ and word~$v$. This finishes the proof of Theorem~\ref{thm:language_equivalence}.
        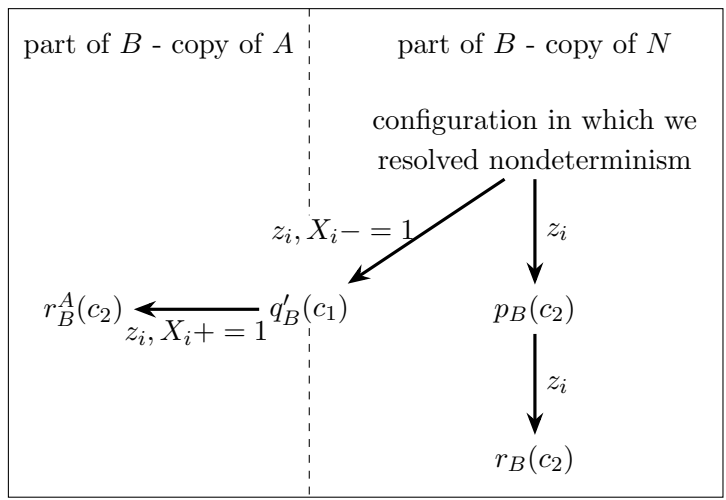
\begin{figure}
    \centering
    \begin{tikzpicture}
        \begin{scope}[initial text = \textcolor{red}{}]
            \node[] (J) at (3,-0.5) {configuration in which we};
            \node[] (B) at (3, -1){resolved nondeterminism};
            \node[] (C) at (0, -3) {$q_B'(c_1)$};
            \node[] (D) at (-3,-3) {$r_B^A(c_2)$};
            \node[] (F) at (3, 0.5) {part of $B$ - copy of $N$};
            \node[] (G) at (-2, 0.5) {part of $B$ - copy of $A$};
            \node[] (H) at (3,-3) {$p_B(c_2)$};
            \node[] (I) at (3, -5) {$r_B(c_2)$};
        \end{scope}
        
        \begin{scope}[>={Stealth[black]},
                      every edge/.style={draw=black,very thick}]
            \draw[] (-4, 1) rectangle (5.5, -5.5);
            \draw[dashed] (0, 1) -- (0, -1.75);
            \draw[dashed] (0, -3.25) -- (0, -5.5);
            \path [->] (B) edge [left] node {$z_i, X_i-=1$} (C);
            \path [->] (C) edge [below] node {$z_i, X_i+=1$} (D);
            \path [->] (B) edge [right] node {$z_i$} (H);
            \path [->] (H) edge [right] node {$z_i$} (I);
        \end{scope}
    \end{tikzpicture}
\caption{Situation in VASS $B$. We have that $q_1'$ comes from $A$ and was created as a copy of state $q_3$ which is a copy of the same state of $N$ as  $q_2'$. Labels $X_i -=1$ and $X_i+=1$ mean decrementing and incrementing $i$-th counter by one, respectively}\label{img:help_2}
\end{figure}
\end{proof}
Theorem \ref{thm:language_equivalence} has multiple corrolaries for various classes of languages, which we present below.
\begin{corollary}
\label{cor:lang_equivalence}
    It is undecidable to determine whether for a given deterministic 2-VASS $A$ and a given history-deterministic 2-VASS $B$ it holds that $L(A) = L(B)$.
\end{corollary}
\begin{proof}
     From Theorem \ref{thm:language_equivalence} we get, that equivalence of trace languages is undecidable. It can be reduced to the equivalence of languages in the coverability semantics by setting the configuration $q(0, 0)$ for every state $q$ of the VASS as an accepting one. Hence the equivalence of languages for the coverability semantics is also undecidable.
\end{proof}
\begin{corollary}\label{cor:lang_equivalence_weak}
     It is undecidable to determine whether for an unambiguous 2-VASS $A$ and a nondeterministic 2-VASS $B$ it holds that $L(A) = L(B)$.
\end{corollary}
\begin{proof}
    We know, that each deterministic VASS is also unambiguous and, that each history-deterministic VASS is also nondeterministic. Hence this corollary follows directly from Corollary \ref{cor:lang_equivalence}.
\end{proof}
\begin{corollary}
    It is undecidable to determine whether for a given unambiguous 2-VASS $A$ and a given nondeterministic 2-VASS $B$ it holds that $L(A) \subseteq L(B)$.
\end{corollary}
\begin{proof}
    In \cite{concur-unambigous} it was shown that containment $L(V_1) \subseteq L(V_2)$ is decidable where $V_1$ is nondeterministic and $V_2$ is unambiguous. Moreover, by Corollary \ref{cor:lang_equivalence_weak} language equivalence problem is undecidable even if one VASS is unambiguous and the second one nondeterministic. Therefore containment $L(A) \subseteq L(B)$ for unambiguous VASS $A$ and nondeterministic VASS $B$ has to be undecidable (otherwise equivalence would be decidable, which is not the case). The same result can be achieved by closer investigation of the proof of Theorem~\ref{thm:language_equivalence}.
\end{proof}
Now we introduce the notion of simulation of one VASS by another one in order to generalise Theorem~\ref{thm:language_equivalence}.
    We say that VASS $V_0$ with an initial configuration $c_0$ simulates VASS $V_1$ with an initial configuration $c_1$ if Duplicator has the winning strategy in the following game:
    \begin{enumerate}
        \item There are two players: Spoiler and Duplicator. Intuitively Spoiler wants to show that $V_0$ does not simulate $V_1$ and Duplicator wants to show that $V_0$ simulates $V_1$.
        \item The initial configuration of Spoiler is $c_1$ and the initial configuration of Duplicator is $c_0$.
        \item Then the following takes place as long as it is possible:
        \begin{enumerate}
            \item If there is transition, which is enabled in the current configuration of Spoiler he loses. Otherwise, he chooses a transition from his current configuration over some letter $a$. He updates his current configuration by firing the chosen transition.
            \item If there is no transition over $a$ from the current configuration of Duplicator he loses. Otherwise, he chooses a transition from his current configuration over letter $a$. He updates his current configuration by firing the chosen transition.
        \end{enumerate}
        \item If the play is infinite then Duplicator wins.
    \end{enumerate}
    Moreover we say that $V_0$ and $V_1$ are in two-sided simulation relation if $V_0$ simulates $V_1$ and $V_1$ simulates $V_0$. 

Observe, that two-sided simulation is an equivalence relation. Two other equivalence relations, which are common in the literature (see \cite{jancar_first}) are trace language equivalence and bisimulation. Bisimulation can be defined similarly to the two-sided simulation, but in each turn, Spoiler chooses in which VASS he wants to make a move and Duplicator has to respond in the other one. For details, we refer to \cite{jancar_first}. It is easy to see, that two-sided simulation is a larger relation than bisimulation (that means if VASSs $A$ and $B$ are in bisimulation relation, they are also in two-sided simulation relation). It is because in bisimulation Spoiler can change VASS during the game, so it wins more often in that case. Moreover, two-sided simulation is smaller than trace language equivalence. This is because if trace languages of VASSs $A$ and $B$ are not equal, without loss of generality, there exists a word $w \in L_T(A)$ for which $w \notin L_T(B)$ holds. Therefore the sequence of transitions reading $w$ in $A$ can be used by Spoiler to show that VASS $B$ can not simulate VASS $A$. 
For any equivalence relation $R$ defined on set of all $2$-VASSs over alphabet $\Sigma$ we say that \naturalrel if the following conditions hold:
 \begin{enumerate}
        \item For each VASSs $A$ and $B$, which are in two-sided simulation relation we have $(A,B) \in R$\label{cond:sim}
        \item $L_T(A) \neq L_T(B)$ implies $(A,B) \notin R$. \label{cond:lang}
\end{enumerate}
\begin{theorem}\label{thm:equivalence}
    For any equivalence relation $R$ defined on set of all $2$-VASSs over alphabet $\Sigma$ such that \naturalrel it is undecidable to determine whether for a given deterministic $2$-VASS $A$ and a given history-deterministic $2$-VASS $B$ it holds that $(A,B) \in R$.
\end{theorem}
For proving Theorem \ref{thm:equivalence} we need the following lemma:
\begin{lemma}\label{lem:equivalence_help}
    For all history-deterministic VASSs $A$ and $B$ it holds that $A$ and $B$ are in two-sided simulation if and only if $L_T(A) = L_T(B)$.
\end{lemma}
\noindent Before proving Lemma \ref{lem:equivalence_help} we show how it implies Theorem \ref{thm:equivalence}.
\begin{proof}[Proof of Lemma \ref{thm:equivalence}]
    Let us fix relation $R$ and VASSs $A$ and $B$. Observe, that if $(A,B) \in R$ then due to Condition \ref{cond:lang} we have $L_T(A) = L_T(B)$. On the other hand if $(A,B) \notin R$ then $A$ and $B$ are not in two-sided simulation, due to Condition \ref{cond:sim},  and hence, because of Lemma \ref{lem:equivalence_help}, $L_T(A) \neq L_T(B)$. Hence $(A,B) \in R \iff L_T(A) = L_T(B)$. Therefore trace language equivalence between deterministic VASS and history-deterministic VASS can be reduced to checking whether they are equivalent with respect to relation $R$. Thus, because of Theorem \ref{thm:language_equivalence}, checking whether $(A,B) \in R$ is undecidable.
\end{proof}
\noindent Now we prove Lemma \ref{lem:equivalence_help}.
\begin{proof}[Proof of Lemma \ref{lem:equivalence_help}]
    It is easy to show, that the fact that $A$ and $B$ are in two-sided simulation implies $L_T(A) = L_T(B)$. Assume, towards contradiction, that $L_T(A) \neq L_T(B)$. Without loss of generality, we can assume, that there exists $w \in L_T(A)$ such that $w \notin L_T(B)$. But then Spoiler can fire sequence of transitions over $w$ in $A$ and Duplicator cannot respond to that in $B$. Therefore $A$ and $B$ are not in two-sided simulation. Contradiction. Hence $L_T(A) = L_T(B)$.

    Now we have to show, that $L_T(A) = L_T(B)$ implies that  $A$ and  $B$ are in two-sided simulation. Therefore it is enough to show, that VASS $A$ simulates VASS $B$ and vice versa.

    We show, that $B$ can simulate $A$. The other case is symmetric.  Observe, that because VASS $B$ is history-deterministic there exists a resolver, which given one letter at the time and history of the run can resolve nondeterminism. We will use this resolver as a strategy for Duplicator. Observe, that after firing some sequence of transitions in $A$ we have then exactly one sequence of transitions to fire over the same word in $B$ given by the resolver. Because languages are equal and resolver correctly resolves nondeterminism in each turn of simulation game Duplicator will be able to respond to an action of Spoiler.
\end{proof}
Observe that Lemma \ref{lem:equivalence_help} cannot be extended to bisimulation, because if one VASS is deterministic, then the bisimulation problem is decidable \cite{jancar_first}. Notice that Theorem \ref{thm:equivalence} can be applied to any equivalence relation between two-sided simulation and trace language equivalence. Notice, that because of Lemma \ref{lem:equivalence_help} deterministic VASS $A$ and a history-deterministic VASS $B$ are either in all of these relations or none. Therefore, to get a wider spectrum of equivalence relations we have to weaken the statement of Theorem \ref{thm:equivalence} by allowing VASS $B$ to be nondeterministic instead of history-deterministic.
\begin{corollary}
    For any equivalence relation $R$ defined on set of all $2$-VASSs over alphabet $\Sigma$ such that \naturalrel it is undecidable to determine whether for a given deterministic $2$-VASS $A$ and a given nondeterministic $2$-VASS $B$ it holds that $(A,B) \in R$.
\end{corollary}

\section{Future research}\label{sec:future}
In this paper we have considered equivalence relations $R$ defined on set of all $2$-VASS such that \naturalrel. The first natural question one can ask is whether our undecidability results hold also when given VASSs have only one counter instead of two. On the other hand one can ask about decidability of bisimulation in one of the following cases:
\begin{enumerate}
    \item history-deterministic vs history-deterministic
    \item history-deterministic vs non-deterministic
\end{enumerate}
Bisimulation is known to be undecidable in general \cite{jancar_first, jancar_second} and decidable if one of the VASS is deterministic\cite{jancar_first}.
\bibliographystyle{elsarticle-num}
\bibliography{main}



\end{document}